\documentclass[letter]{aa} 
\usepackage{graphicx}
\usepackage[varg]{txfonts}
\usepackage{rotating}
\usepackage{lscape}
\usepackage{natbib}
\newcommand{\asec}{$^{\prime\prime}$}

\def\H{N$_{2}$H$^{+}$}
\def\D{N$_{2}$D$^{+}$}

\def\AMM{NH$_3$}

\def\kms{\mbox{km~s$^{-1}$}}

\def\cmq{cm$^{-2}$}

\def\Vlsr{$V_{\rm LSR}$}
\def\Dfrac{$D_{\rm frac}$}
\def\Tex{\mbox{$T_{\rm ex}$}}

\def\Lsun{\mbox{$L_{\odot}$}}

\def\Tk{\mbox{$T_{\rm k}$}}

\begin{document}
\title{Deuteration as an evolutionary tracer in massive-star formation
\thanks{Based on observations carried out with the IRAM 30m telescope.
IRAM is supported by INSU/CNRS (France), MPG (Germany), and IGN (Spain).}}

   \author{F. Fontani
          \inst{1,2}
          \and
          Aina Palau\inst{3} \and  P. Caselli \inst{4}
          \and \'A. S\'anchez-Monge \inst{5}
          \and M. J. Butler \inst{6} \and J. C. Tan \inst{6,7}
          \and I. Jim\'enez-Serra \inst{8}
           \and G. Busquet \inst{5,9}
          \and S. Leurini \inst{10}
        \and M. Audard \inst{11,12}
          }

\institute{ESO, Karl Schwarzschild Str. 2, 85748 Garching bei Munchen, Germany
        \and
           Institut de Radio-Astronomie Millim\'etrique, 300 rue de la Piscine, 38406 Saint Martin d'H\`eres, France \\
           \email{fontani@iram.fr}
           \and 
           Institut de Ci\`encies de l'Espai (CSIC-IEEC), Campus UAB-Facultat de Ci\`encies, Torre C5-parell 2, E-08193 Bellaterra, Spain
            \and
           School of Physics and Astronomy, E.C. Stoner Building, University of Leeds, Leeds LS2 9JT 
            \and
           Departament d'Astronomia i Meteorologia (IEEC-UB), Institut de Ci\`encies del Cosmos, Universitat de Barcelona, Mart\'i i Franqu\`es, 1, E-08028 Barcelona, Spain
           \and 
           Department of Astronomy, University of Florida, Gainesville, FL 32611, USA
           \and 
           Department of Physics, University of Florida, Gainesville, FL 32611, USA
           \and
           Harvard-Smithonian Center for Astrophysics, 60 Garden St., Cambridge MA 02138, USA
           \and
           INAF-Istituto di Fisica dello Spazio Interplanetario, Via Fosso del Cavaliere 100, 00133, Roma, Italy
            \and
           Max-Planck-Institut fur Radioastronomie, Auf dem Hugel 69, 53121, Bonn, Germany
           \and
           ISDC Data Center for Astrophysics, University of Geneva, Ch. d'Ecogia 16, 1290 Versoix, Switzerland 
           \and
           Geneva Observatory, University of Geneva, ch. des Maillettes 51, 1290 Versoix, Switzerland
           } 
\date{Received date; accepted date}

\titlerunning{Deuteration in massive-star formation}
\authorrunning{Fontani et al.}

 
 \abstract{Theory predicts, and observations confirm, that the column density
 ratio of a molecule containing D to its counterpart containing H can be used as an 
 evolutionary tracer in the low-mass star formation process.}
 {Since it remains unclear if the high-mass star formation process is a scaled-up
 version of the low-mass one, we investigated whether the relation between deuteration
 and evolution can be applied to the high-mass regime.}
 {With the IRAM-30m telescope, we observed rotational transitions of \D\ and
 \H\ and derived the deuterated fraction in 27 cores within massive star-forming
 regions understood to represent different evolutionary stages of the massive-star formation process.}
 {Our results clearly indicate that the abundance of \D\ is higher at the pre--stellar/cluster
 stage, then drops during the formation of the protostellar object(s) as in the low-mass regime,
 remaining relatively constant during the ultra-compact HII region phase. 
 The objects with the highest fractional abundance of \D\ are starless cores with properties
 very similar to typical pre--stellar cores of lower mass. The abundance of \D\ is lower in objects 
 with higher gas temperatures as in the low-mass case but does not seem to depend on gas turbulence. }
 {Our results indicate that the \D -to-\H\ column density ratio can be used as an evolutionary 
 indicator both in low- and high-mass star formation, and that the physical conditions that
 influence the abundance of deuterated species likely evolve similarly along
the processes that lead to the formation of both low- and high-mass stars.}
  

\maketitle

 \maketitle
%

 \section{Introduction}
 \label{intro}
 
 The study of deuterated molecules is an extremely useful probe of the physical conditions in 
 star-forming regions. Deuterated species are readily produced in molecular
environments characterised by low temperatures ($T\leq 20$ K) and CO depletion (Millar et al.~1989).
These physical/chemical properties are commonly observed in low-mass pre--stellar cores (starless cores
on the verge of forming stars), where the {\it deuterated fraction} (hereafter \Dfrac ) of non-depleted 
molecules, defined as the column density ratio of one 
species containing deuterium to its counterpart containing hydrogen,
is orders of magnitude larger than the [D/H] interstellar abundance (of the order
of $\sim 10^{-5}$, Oliveira et al.~2003). 
Caselli~(2002a) found a theoretical relation between \Dfrac\ and the core evolution in the low-mass case. 
This relation predicts that \Dfrac\ increases when the starless core 
evolves towards the onset of gravitational collapse because, as the core density profile becomes
more and more centrally peaked, freeze-out of CO increases in the core centre
and hence the abundance of deuterated molecules is greatly enhanced.
When the young stellar object formed at the core centre begins to heat its surroundings, the 
CO evaporated from dust grains starts to destroy the deuterated species and
\Dfrac\ decreases. 
Observations of both starless cores and cores with already formed protostars confirm the
theoretical predictions in the low-mass regime: the pre--stellar cores closest to gravitational 
collapse have the highest \Dfrac\ (Crapsi et al.~2005), while \Dfrac\ is lower in cores 
associated with Class 0/I protostars, and the coldest (i.e. the youngest) objects possess 
the largest \Dfrac , again in agreement with the predictions of chemical 
models (Emprechtinger et al.~2009, Friesen et al.~\citeyear{friesen}).
On the basis of these results, \Dfrac\ can be considered as an evolutionary
tracer of the low-mass star formation process before and after the formation of the protostellar object.

Can this result be applied to the high-mass regime? This question is difficult to answer
because the massive-star formation process is still not well-understood: large distances 
($\geq 1$ kpc), high extinction and clustered environments
make observations of the process challenging (Beuther et al~\citeyear{beuther07},
Zinnecker \& Yorke~2007).
Observationally, the study of Pillai et al.~(2007), performed with the Effelsberg
and IRAM-30m telescopes, measured high values of \Dfrac\ ($\sim 0.2$) from 
deuterated ammonia in infrared dark clouds, which are understood to
represent the earliest stages of massive star and stellar cluster formation.
In more evolved objects, from IRAM-30m observations, Fontani et al.~(2006) 
measured smaller values of \Dfrac\ ($\sim 10^{-2}$) from the ratio \D /\H ,
which are nevertheless much larger than the D/H interstellar abundance.
Despite these efforts, 
no systematic study of the [D/H] ratio across all stages of high-mass star formation
has yet been carried out.

In this letter, we present the first study of the relation between deuterated fraction
and evolution in a statistically significant sample of cores embedded in high-mass star 
forming regions spanning a wide range of evolutionary stages, 
from high-mass starless core candidates (HMSCs) to high-mass protostellar objects (HMPOs) and 
ultracompact (UC) HII regions (for a definition of these stages see e.g. Beuther
et al~\citeyear{beuther07}). 
This goal was achieved by observing rotational transitions of \H\ and \D\ with
the IRAM-30m telescope. We chose these species because \D\ can be formed from \H\
only in the gas phase, tracing cold and dense regions 
more precisely than deuterated \AMM , which can also be formed on dust grains 
(e.g. Aikawa et al.~\citeyear{aikawa}) and then evaporates 
by heating from nearby active star-formation. 
Even though the observations are obtained with low
angular resolution, the objects observed in the survey were carefully selected to
limit as much as possible any emission arising from adjacent 
objects.

\section{Source selection and observations}
\label{obs}

The source list is in Table~\ref{tab_sources}, where we give the source
coordinates, the distance, the bolometric luminosity, and the reference 
papers. We observed 27 molecular cores divided into: ten HMSCs, 
ten HMPOs, and seven UC HII regions. 
The source coordinates were centred 
towards either (interferometric) infrared/millimeter/centimeter continuum peaks or 
high-density gas tracer peaks (\AMM\ with VLA, \H\ with CARMA or PdBI) identified 
in images with angular resolutions comparable to or better than 6\arcsec , 
either from the literature or from observations not yet published. 
In general, we rejected objects whose emission peaks were
separated by less than $\sim 8$\asec\ 
from another peak of a dense molecular gas tracer.
This selection criterion was adopted to avoid or limit as much as possible
the presence of multiple cores within the IRAM-30m beam(s).
The evolutionary stage of each source was established based on a 
collection of evidence: HMSCs are massive cores embedded in
infrared dark-clouds or other massive star forming regions not associated 
with indicators of ongoing star formation (embedded 
infrared sources, outflows, masers); HMPOs are
associated with interferometric powerful outflows, and/or infrared sources, and/or 
faint ($S_{\nu}$ at 3.6~cm $< 1 $mJy) radio continuum emission likely tracing a radio-jet; 
and UC HIIs must be associated with a stronger radio-continuum ($S_{\nu}$  at 3.6~cm $\geq$ 1 mJy) 
that probably traces gas photoionised by a young massive star.
We did not include evolved HII regions that have already dissipated the associated molecular core.
We also limited the sample to sources at distances of less than $\sim 5$ kpc.
We stress that the three categories must be regarded with
caution because it can be difficult to determine the relative evolutionary stage. 
This caveat applies especially to HMPOs and UC HII regions,
whose evolutionary distinction is not always a clear cut (see e.g. Beuther et al.~\citeyear{beuther07}).
Among the HMSCs, three sources (AFGL5142- EC, 05358-mm3, and I22134-G)
have been defined as "warm" in Table~\ref{tab_sources}:
we explain the peculiarity of these sources in Sect.~\ref{res}.
The observations of the 27 cores listed in Table~\ref{tab_sources}
were carried out with the IRAM-30m telescope in two main observing runs
(February 2 to 4, June 19 to 21, 2010), and several additional hours
allocated during three Herapool weeks (December 2009, January 2010, 
and November 2010).
We observed the \H\ (3--2), \H\ (1--0), and \D\ (2--1) transitions. 
The main observational parameters of these lines are given in 
Table~\ref{tab_mol}.
The observations were made in wobbler--switching mode. Pointing 
was checked every hour. The data were calibrated with the chopper wheel 
technique (see Kutner \& Ulich~\citeyear{kutner}), with a calibration
uncertainty of $\sim 20 - 30\%$. The spectra were obtained in
antenna temperature units, $T_{\rm A}^{*}$, and then converted to
main beam brightness temperature, $T_{\rm MB}$, via the relation
$T_{\rm A}^{*}=T_{\rm MB}\,\eta_{\rm MB}$, where
$\eta_{\rm MB}=B_{\rm eff}/F_{\rm eff}$ is 0.74 for \D\ (2--1), 
0.53 for \H\ (3--2) and 0.88 for \H\ (1--0) lines, respectively. 
All observed transitions possess hyperfine structure. To take this into account, 
we fitted the lines using
METHOD HFS of the CLASS program, which is part of the GILDAS 
software\footnote{The GILDAS software is available at http://www.iram.fr/IRAMFR/GILDAS}
developed at the IRAM and the Observatoire de Grenoble. This method assumes
that all the hyperfine components have the same excitation temperature
and width, and that their separation is fixed to the laboratory value.
The method also provides an estimate of the optical depth of
the line, based on the intensity ratio of the different hyperfine
components. For the faintest \D\ lines, for which the hfs method gives 
poor results, the lines were fitted assuming a Gaussian shape.

\section{Results and discussion: is deuteration an evolutionary indicator of massive star formation?}
\label{res}

The spectra of \D\ (2--1) and \H\ (3--2) for all sources detected in \D\ are
shown in Figures~\ref{spectra1} -- \ref{spectra6}. We detected
\H\ (3--2) emission in all sources. We also found a remarkably high 
detection rate in the \D\ (2--1) line: 100$\%$ in HMSCs, 64$\%$ in HMPOs,
and 100$\%$ in UC HII regions.
Such a high detection rate indicates that deuterated gas is present 
at every stage of the massive star and star cluster formation process, even in the 
surroundings of UC HII regions where the gas is expected to be hotter
and more chemically evolved.
Even though for 12 sources we also observed the \H\ (1--0) transition,
we always computed the column density of \H\ and the deuterated fraction 
from the (3--2) line given its smaller telescope beam, to limit the contribution of
nearby sources as much as possible. An overall presentation of the data obtained, 
and a deeper analysis of all physical parameters, will be given in a 
forthcoming paper. 
We derived the \H\ and \D\ column densities, $N({\rm N_2H^+})$ and
$N({\rm N_2D^+})$, from the line integrated intensity following the method described in the 
appendix of Caselli et al.~(\citeyear{casellib}). Thanks to the
selection criteria for our sources, for which interferometric maps of
dense gas are available for most of the regions, a first estimate of
the filling factor could be computed. However, because maps of the 
two transitions used to derive \Dfrac\  have not yet been performed (except for I22134-VLA1),
the source size was determined from interferometric measurements
of \AMM (2,2). This assumption seems reasonable because this line 
traces gas with physical conditions similar
to those of \H\ (3--2) and \D\ (2--1). To take into account the
possible effects of the evolutionary stage on the source size, we also computed an 
average diameter for each evolutionary group. This turns out to 
be: 6.5\arcsec\ for HMSCs, 4.1\arcsec\ for HMPOs, and 5.5\arcsec\ for UC HIIs 
(Busquet~2010, Busquet et al.~2011, S\'anchez-Monge~2011, 
Palau et al.~2007, 2010). We stress that these angular diameters are 
consistent with the (few) \H\ and \D\ interferometric observations 
published to date (e.g. see the case of IRAS 05345+3157, Fontani et al. 2008).
The \H\ and \D\ column densities, their ratio (\Dfrac ), as well
as the line parameters used in the derivation
of the column densities, are listed in Table~\ref{tab_res}. 

The method assumes a constant excitation temperature, \Tex .
For the \H\ lines, \Tex\ was derived directly from the parameters given
by the hyperfine fitting procedure corrected
for the filling factor (see the CLASS user manual for
details: http://iram.fr/IRAMFR/GILDAS/doc/html/class-html/class.html).
The procedure, however, cannot provide good estimates for 
optically thin transitions or transitions with opacity ($\tau$) not well-constrained (e.g.
with relative uncertainty larger than $30 \%$). For these, we
were obliged to assume a value for \Tex\ (for details, see the notes of Table~\ref{tab_res}). 
For the \D\ (2--1) lines we were unable to derive \Tex\ from the fitting procedure for almost all
sources because $\tau$ is either too small or too uncertain.
In 3 cases only was the optical depth of the \D\ (2--1) transition well-determined, and so is \Tex : 
in two of these objects we found a close agreement between the estimates derived from 
the \D\ (2--1) and the \H\ (3--2) transitions. Therefore, the \D\ 
column density of each source was computed assuming the same \Tex\ as for \H .
Since \D\ (2--1) and \H\ (3--2) have similar critical densities and we measure similar
\Tex\ for both transitions, the two lines approximately trace similar material, so that
computing \Dfrac\ using them is a reasonable approach.
The \H\ column densities are on average of the order of $10^{13 - 14}$ \cmq , and
the \D\ column densities are of order $10^{12 - 13}$ \cmq .
Both values are consistent with similar observations towards massive star forming
regions (e.g. Fontani et al.~\citeyear{fontani06}). The measured \Tex\ 
corrected for filling factor are between $\sim 7$ and $\sim 50$ K and agree, on average,
with the kinetic temperatures measured from ammonia, except for the colder HMSCs
for which they are a factor of $\sim 2$ lower.

\begin{figure}
\centerline{\includegraphics[angle=-90,width=9cm]{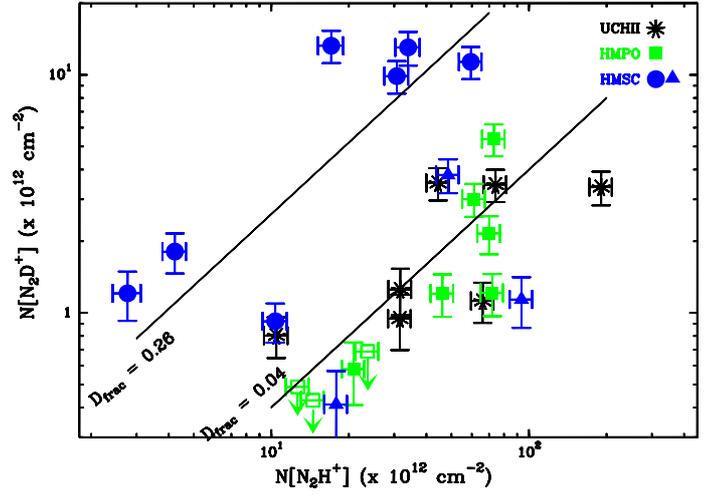}}
 \caption{\D\ column density versus \H\ column density. Blue symbols correspond to
 HMSCs (triangles: "warm" cores, see text); green
 squares show HMPOs (open squares are upper limits); black asterisks
 correspond to UC HII regions. The two lines indicate the average values of
 \Dfrac\ for the HMSC group (i.e. 0.26) and that of both the HMPO and 
 UC HII groups (i.e.~0.04). }
\label{fig_dfrac}
\end{figure}


\begin{figure}
\centerline{\includegraphics[angle=-90,width=9cm]{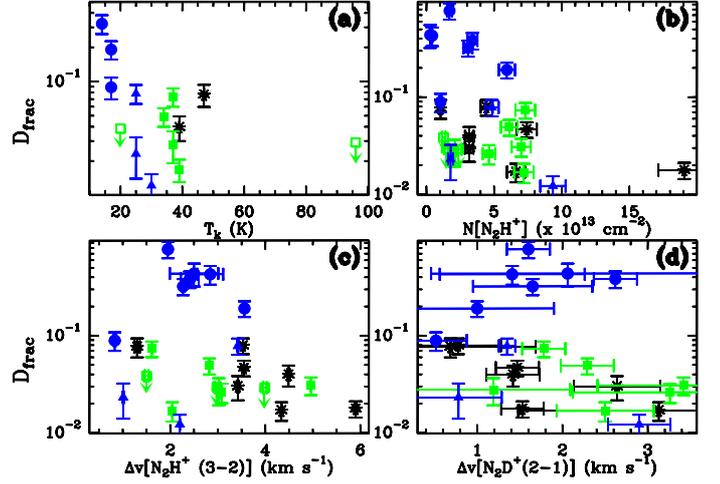}}%
 \caption{Deuterated fraction, \Dfrac = $N$(\D )$/N$(\H ), as a function of several parameters:
  kinetic temperature (a), N(\H ) (b), \H\ (3--2) line width (c) and \D\ (2--1) line width (d).
 The symbols have the same meaning as in Fig.~\ref{fig_dfrac}. For some sources, the
 errorbars are not visible because they are smaller than the symbol size. In panel (a), only the
 sources with temperature derived from VLA ammonia observations are plotted.
 }
 \label{fig_dfrac_etal}
 \end{figure}

The deuterated fraction for the three
evolutionary groups is shown in Fig.~\ref{fig_dfrac}, where we plot
$N$(\D ) against $N$(\H ). There is a statistically significant
separation between the HMSC group, which has the highest average \Dfrac\ 
(mean value $\sim 0.26$, $\sigma = 0.22$), and the HMPOs and UC HII groups, which 
have similar average deuterated fraction: mean \Dfrac $= 0.037$ ($\sigma = 0.017$) for HMPOs,
and mean \Dfrac = 0.044 ($\sigma = 0.024$) for UC HII regions. Both are about an order of
magnitude smaller than that associated with HMSCs.
A closer inspection of the data using the Kolmogorov-Smirnov statistical test
shows that the separation in \Dfrac\ between the HMSC group and
that including both HMPOs and UC HII regions is indeed statistically
significant: the test shows that the probability of the
distributions being the same is very low ($P\sim 0.004$). This is
strong evidence that the two groups differ statistically. 
Therefore, massive cores without stars have larger abundances of \D\
than cores with already formed massive (proto-)stars or proto-clusters.
The abundance of \D , however, seems to remain constant, within
the uncertainties, after the formation of the protostellar object until the UC HII region 
phase.
That \Dfrac\ is of the order of $\sim 0.2- 0.3$, on average, in HMSCs, and then 
drops by an order of magnitude after the onset of star formation, 
indicates that the physical conditions acting on the abundance
of deuterated species (i.e. density and temperature) evolve similarly along both
the low- and high-mass star formation processes (see e.g.~Crapsi et
al.~\citeyear{crapsi} and Emprechtinger et al.~\citeyear{emprechtinger}). 
%

Another interesting aspect emerging from Fig.~\ref{fig_dfrac} is that the
three HMSCs defined as "warm" in Table~\ref{tab_sources}
(AFGL5142-EC, 05358-mm3, and I22134-G, marked as triangles in
the figure) have \Dfrac\ almost an order of magnitude smaller
than the others. These differ from the rest of the sub-sample of HMSCs 
because they have temperatures \Tk\ $> 20$ K (see Table~\ref{tab_res} and panel (a) in 
Fig.~\ref{fig_dfrac_etal}). High angular resolution studies indicate that they 
could be externally heated
(Zhang et al.~\citeyear{zhang02}, Busquet~\citeyear{busquetphd},
S\'anchez-Monge~\citeyear{sanchez11}), so that they are likely to be perturbed by nearby
star formation and we expect their properties to be different from those
of the other, more quiescent cores.
An anticorrelation between \Dfrac\ and the distance to heating sources such as
embedded protostars was found in the cluster-forming Ophiuchus-B 
clump by Friesen et al~(\citeyear{friesen}). Our study tends to confirm
the Friesen et al.'s finding, even
though the poor statistics does not allow us to drive firm conclusions.
We also point out that the four cores selected from the Butler \& Tan~(2009)
work (G034-G2, G034-F1, G034-F2, G028-C1) have the highest values of
all measured \Dfrac\ and lie in infrared-dark regions, away from active star formation.
These four cores are hence very similar to the prototype low-mass 
'pre-stellar cores' (e.g. L1544, L694--2, see Crapsi et al.~2005) and we propose
that these are good 'massive pre--stellar core' candidates.

In Fig.~\ref{fig_dfrac_etal}, we plot \Dfrac\ as a function of 
several parameters: the kinetic temperature, the \H\ column
density, and the
line widths derived from both \H\ and \D . 
To search for possible (anti-)correlations between these parameters, we performed
two statistical tests: the Kendall's $\tau$ and the Spearman's $\rho$ rank correlation tests
\footnote{(http://www.statsoft.com/textbook/nonparametric-statistics/ )}. 
For \Tk , the tests were applied to all sources in our survey with 
gas temperature derived from VLA interferometric ammonia observations
(see Table~\ref{tab_res}).
As it can be inferred from panel (a) of Fig.~\ref{fig_dfrac_etal}, 
\Dfrac\ and \Tk\ are slightly anti-correlated 
($\tau = -0.38$, $\rho$= $-0.50$), and
 \Dfrac\ is also anti-correlated with the \H\ column density 
($\tau =-0.43$, $\rho$= $-0.60$, panel (b) in Fig.~\ref{fig_dfrac_etal}). 
We also find a very faint 
anticorrelation between \Dfrac\ and the \H\ line width ($\tau =-0.17$, $\rho=-0.23$) and 
between \Dfrac\ and the \D\ line width ($\tau=-0.25$, $\rho=-0.35$)
(panels (c) and (d) in in Fig.~\ref{fig_dfrac_etal}, respectively).
In particular, this latter is difficult to trust being affected by large uncertainties 
in the \D\ line widths.
Emprechtinger et al.~(\citeyear{emprechtinger})
suggested that in low-mass star forming cores the deuteration is higher in colder and 
more quiescent cores, according to the predictions of theoretical models. 
A similar trend was found also in a small sample of seven massive star-forming clumps by 
Fontani et al.~(\citeyear{fontani06}) including both HMPOs and UC HII regions but
not HMSCs.
That the warmer sources have lower \Dfrac\ is not surprising and can be explained
by the CO freeze-out and the chemical reactions leading to the enhancement
of deuterium abundance being strongly depressed when the temperature 
increases (Caselli et al.~\citeyear{caselli08}). The lack of correlation 
between deuterated fraction and line widths tells us that the deuterium fractionation 
process is independent of the gas turbulence.
This result agrees with high-angular resolution observations
of cluster-forming regions (Fontani et al.
~\citeyear{fontani09}, Busquet et al.~\citeyear{busquet}), but given 
the large uncertainties (especially on the \D\ line widths), 
the conclusions must be interpreted with caution.
We speculate that the anticorrelation between \Dfrac\ and $N$(\H ) 
could indicate that, assuming that \Dfrac\ decreases in the protostellar
phase, the \H\ column density increases during the younger and most
embedded period of the protostellar phase, as suggested by Busquet~(2010)
for a different sample of sources.

In summary, our findings indicate that the physical conditions acting on the abundance
of deuterated species (i.e. density and temperature) evolve similarly during both
the low- and high-mass star formation process.
To confirm this, several questions however need to be answered: 
in HMSCs, do the \D\ and \H\ emission peak at dust emission peak as in low-mass 
 pre--stellar cores? What is the nature of the \D\ emission 
 in evolved objects (HMPOs and UC HII regions)? Is the emission extended or fragmented 
 into several condensations (as found in the few massive star forming regions observed 
 with interferometers)? To answer these questions, higher 
 angular resolution observations are necessary.
 On the theoretical side, we also need to investigate this proposed evolutionary sequence
using astrochemical models.

{\it Acknowledgments.} 
We are grateful to the IRAM-30m telescope staff for their help during the observations.
Many thanks to the anonymous Referee for his/her comments that significantly 
improved the work. FF has received funding from
the European Community's Seventh Framework Programme (FP7/2007--2013)
under grant agreement No. 229517.
AP, AS-M and GB are supported by the Spanish MICINN grant AYA2008-06189-C03
(co-funded with FEDER funds). AP is supported by JAEDoc CSIC fellowship
co-funded with the European Social Fund.
GB is funded by the Italian Space Agency (ASI) with contract ASI-I/005/07/1.
MA acknowledges support from the Swiss National Science Foundation
(grants PP002-110504 and PP00P2-130188)

{}

\clearpage

\renewcommand{\thetable}{A-\arabic{table}}
\setcounter{table}{0}
\section*{Appendix A: Tables}
\label{appa}

Table~\ref{tab_sources} contains the list of the observed sources selected as
explained in Sect.~2 of the main body text, and give some information 
extracted from the literature about the star forming regions in which the sources
lie. Table~\ref{tab_mol} presents the observed transitions and some main
technical observational parameters. Table~\ref{tab_res} shows the results of 
the fitting procedure to the \D\ (2--1) and \H\ (3--2) lines (see Sect.~2 of the main body text)
of all sources, and the physical parameters derived from these results,
namely the \H\ and \D\ column densities and their ratio, \Dfrac .
Other parameters discussed in Sect.~3 of the main body text are
also listed.

\addtocounter{table}{0}

\begin{table*}
\begin{center}
\caption[] {list of the observed sources. Col.~4 shows the velocity at which
we centred the spectra, corresponding to the systemic velocity. Cols.~5 and 6 give 
the source distance and bolometric luminosity (respectively) of the associated star 
forming region. This latter is a very rough first approximation of the core
luminosity because it is based on
infrared measurements having poor angular resolution.
We adopt as source names those used in the reference papers listed in Col.~7.}
\label{tab_sources}
\normalsize
\begin{tabular}{ccccccc}
\hline \hline
source& RA(J2000) & Dec(J2000) & \Vlsr\ & $d$ & $L_{\rm bol}$ & Ref. \\
    & h m s & $o$ $\prime$  $\prime\prime$ & \kms\   & kpc & \Lsun\ & \\
\cline{1-7}
\multicolumn{7}{c}{HMSC} \\
\cline{1-7}
I00117-MM2\tablefootmark{a} & 00:14:26.3	& +64:28:28 & $-36.3$ & 1.8  & $10^{3.1}$ & (1) \\
AFGL5142-EC\tablefootmark{b}\tablefootmark{w}  & 05:30:48.7    &  +33:47:53 & $-3.9$ & 1.8 & $10^{3.6}$ & (2) \\
05358-mm3\tablefootmark{b} \tablefootmark{w}  & 05:39:12.5 & +35:45:55 & $-17.6$ & 1.8 & $10^{3.8}$ & (3,11) \\ 
G034-G2(MM2)\tablefootmark{a} & 18:56:50.0 & +01:23:08 &  $+43.6$ & 2.9 & $10^{1.6}$\tablefootmark{r} & (4)  \\
G034-F1(MM8)\tablefootmark{a} & 18:53:19.1  & +01:26:53 & $+57.7$  & 3.7 & $10^{1.9}$\tablefootmark{r}   & (4)  \\
G034-F2(MM7)\tablefootmark{a} & 18:53:16.5  & +01:26:10 & $+57.7$  & 3.7 & -- & (4) \\
G028-C1(MM9) \tablefootmark{a} & 18:42:46.9  & $-$04:04:08 & $+78.3$  & 5.0  & -- & (4) \\
I20293-WC\tablefootmark{a} & 20:31:10.7  &	 +40:03:28 & $+6.3$ & 2.0 & $10^{3.6}$ & (5,6) \\
I22134-G\tablefootmark{b} \tablefootmark{w}  & 22:15:10.5  &   +58:48:59 & $-18.3$  & 2.6 & $10^{4.1}$ & (7) \\
I22134-B\tablefootmark{b}  & 22:15:05.8 &	+58:48:59 & $-18.3$ & 2.6  & $10^{4.1}$ & (7) \\
\cline{1-7}
\multicolumn{7}{c}{HMPO}   \\
\cline{1-7}
I00117-MM1\tablefootmark{a} & 00:14:26.1	& +64:28:44 & $-36.3$  & 1.8 & $10^{3.1}$ & (1) \\
I04579-VLA1\tablefootmark{c}   & 05:01:39.9 &	+47:07:21 & $-17.0$  & 2.5 & $10^{3.6}$ &  (8) \\
AFGL5142-MM\tablefootmark{b}  & 05:30:48.0	& +33:47:54 &  $-3.9$ & 1.8  & $10^{3.6}$ & (2) \\
05358-mm1\tablefootmark{b}  & 05:39:13.1 & +35:45:51 & $-17.6$ & 1.8  & $10^{3.8}$ & (3) \\
18089--1732\tablefootmark{b}  & 18:11:51.4 & $-$17:31:28 & $+32.7$  & 3.6 & $10^{4.5}$ & (9) \\
18517+0437\tablefootmark{b}  & 18:54:14.2 & +04:41:41 & $+43.7$  & 2.9 & $10^{4.1}$ & (10) \\
G75-core\tablefootmark{a} & 20:21:44.0 &	+37:26:38 & $+0.2$  & 3.8 & $10^{4.8}$ & (11,12) \\
I20293-MM1\tablefootmark{a} & 20:31:12.8 &	 +40:03:23 & $+6.3$ & 2.0 & $10^{3.6}$ & (5) \\
I21307\tablefootmark{a} & 21:32:30.6  &    +51:02:16  & $-46.7$ & 3.2 & $10^{3.6}$ & (13) \\ 
I23385\tablefootmark{a} & 23:40:54.5 &      +61:10:28 & $-50.5$   & 4.9 & $10^{4.2}$ & (14) \\
\cline{1-7}
\multicolumn{7}{c}{UC HII}   \\
\cline{1-7}
G5.89--0.39\tablefootmark{b}  & 18:00:30.5    &    $-$24:04:01 & $+9.0$ & 1.28 & $10^{5.1}$ & (15,16) \\
I19035-VLA1\tablefootmark{b}  & 19:06:01.5 &	+06:46:35 & $+32.4$  & 2.2 & $10^{3.9}$ & (11) \\
19410+2336\tablefootmark{a} & 19:43:11.4 &    +23:44:06 & $+22.4$ & 2.1 & $10^{4.0}$ & (17) \\
ON1\tablefootmark{a} & 20:10:09.1  &     +31:31:36 & $+12.0$  & 2.5 & $10^{4.3}$ & (18,19) \\
I22134-VLA1\tablefootmark{a} & 22:15:09.2 &	+58:49:08 & $-18.3$ & 2.6 & $10^{4.1}$ & (11) \\
23033+5951\tablefootmark{a} & 23:05:24.6 & +60:08:09 & $-53.0$  & 3.5 & $10^{4.0}$ & (17) \\
NGC7538-IRS9\tablefootmark{a}  & 23:14:01.8   &   +61:27:20 & $-57.0$  & 2.8 & $10^{4.6}$ & (8) \\
\hline
\end{tabular}
\end{center}
\tablefoot{
\tablefoottext{a}{Observed in \H\ (3--2) and \D\ (2--1);}
\tablefoottext{b}{Observed in \H\ (1--0), \H\ (3--2), and \D\ (2--1);}
\tablefoottext{c}{Observed in \H\ (1--0) and \D\ (2--1);}
\tablefoottext{w}{"warm" HMSCs;}
\tablefoottext{r}{Luminosity of the core and not of the whole associated star-forming region (Rathborne et al.~\citeyear{rathborne});}
\tablefoottext{1}{Palau et al.~(\citeyear{palau10})}
\tablefoottext{2}{Busquet et al.~(\citeyear{busquet11})}
\tablefoottext{3}{Beuther et al.~(\citeyear{beuther07b})}
\tablefoottext{4}{Butler \& Tan~(\citeyear{bet})}
\tablefoottext{5}{Palau et al.~(\citeyear{palau07})}
\tablefoottext{6}{Busquet et al.~(\citeyear{busquet})}
\tablefoottext{7}{Busquet~(\citeyear{busquetphd})}
\tablefoottext{8}{S\'anchez-Monge et al.~(\citeyear{sanchez})}
\tablefoottext{9}{Beuther et al.~(\citeyear{beuther04})}
\tablefoottext{10}{Schnee \& Carpenter~(\citeyear{schnee})}
\tablefoottext{11}{S\'anchez-Monge~(\citeyear{sanchez11})}
\tablefoottext{12}{Ando et al.~(\citeyear{ando})}
\tablefoottext{13}{Fontani et al.~(\citeyear{fonta04a})}
\tablefoottext{14}{Fontani et al.~(\citeyear{fonta04b})}
\tablefoottext{15}{Hunter et al.~(\citeyear{hunter})}
\tablefoottext{16}{Motogi et al.~(\citeyear{motogi})}
\tablefoottext{17}{Beuther et al.~(\citeyear{beuther02})}
\tablefoottext{18}{Su et al.~(\citeyear{su})}
\tablefoottext{19}{Nagayama et al.~(\citeyear{nagayama})}
}
\end{table*}

\addtocounter{table}{0}

\normalsize
\begin{table*}
\begin{center}
\caption[] {Observed transitions and technical parameters}
\label{tab_mol}
\begin{tabular}{ccccc}
\hline \hline
molecular  & frequency & HPBW & $\Delta v$\tablefootmark{a} & Bandwidth\tablefootmark{a} \\
 transition  & (GHz) & (\asec ) & (\kms ) & (\kms ) \\
\hline
\H\ (1$-$0)  & 93.17376\tablefootmark{b} & 26 & $0.126$ & $ 230$ \\
\H\ (3$-$2)   & 279.51186\tablefootmark{c} & 9 & $0.042$ & $77$ \\
\D\ (2$-$1)  & 154.21718\tablefootmark{d} & 16 & $0.076$ & $139$ \\
\hline
\end{tabular}
\tablefoot{
\tablefoottext{a}{Resolution ($\Delta v$) and bandwidth of the spectrometer used (VESPA).}
\tablefoottext{b}{frequency of the main hyperfine component ($F_{1} F = 2\;3\rightarrow 1\;2$ , Pagani et al.~\citeyear{pagani})}
\tablefoottext{c}{frequency of the $F_{1} F = 4\;5\rightarrow 3\;4$ hyperfine component, having a relative intensity of 
$17.46\%$. (Crapsi et al.~\citeyear{crapsi})}
\tablefoottext{d}{frequency of the main hyperfine component ($F_{1} F = 2\;3\rightarrow 1\;2$, Pagani et al.~\citeyear{pagani})}
}
\end{center}
\end{table*}

\clearpage

\addtocounter{table}{0}

\begin{sidewaystable*}
\normalsize
\caption{Observational and derived line parameters: Cols. 2--6 list integrated intensity over all the hyperfine components
($\int T_{\rm MB}{\rm d}v$), peak velocity ($V_{\rm LSR}$),
full width at half maximum (FWHM) corrected for hyperfine splitting, opacity ($\tau$) and excitation temperature (\Tex ), from \H\ (3--2); 
Cols.~7, 8 and 9 list $\int T_{\rm MB}{\rm d}v$, $V_{\rm LSR}$ and FWHM for the \D\ (2--1) lines. In cols~10 and 11 we give the \H\ and \D\
column densities, respectively, and their ratio (\Dfrac ) is given in Col.~12. 
Col.~13 shows the kinetic temperature (\Tk ) measured from ammonia observations.
Uncertainties are given in parentheses.}
\label{tab_res}
\begin{center}
\scriptsize
\begin{tabular}{cccccccccccccccc}
\hline \hline
 &   & \multicolumn{5}{c}{ \H\ (3--2) } & &\multicolumn{3}{c}{ \D\ (2--1) } & & & & \\
\cline{3-7} \cline{9-11}
 & source     & $\int T_{\rm MB}{\rm d}v$ & $V_{\rm LSR}$ &  FWHM & $\tau$ & \Tex\ \tablefootmark{\dagger} & & $\int T_{\rm MB}{\rm d}v$ & $V_{\rm LSR}$ & FWHM  & N(\H ) & N(\D ) & \Dfrac & \Tk\ \tablefootmark{\dagger\dagger} \\
    &   & (K \kms ) & (\kms ) & (\kms ) & & (K) & & (K \kms ) & (\kms ) & (\kms ) & ($\times 10^{13}$ \cmq ) & ($\times 10^{12}$ \cmq ) & & (K) \\
\cline{1-15}
 HMSCs & I00117-MM2  & 2.9(0.1) &  --50.51(0.01) & 2.27(0.04) & $<$0.1 & 7\tablefootmark{a} &                     & 0.48(0.02) & --50(1) & 1.65(0.7) & 3.1(0.3)          &  10(1) &   0.32(0.06) & 14\tablefootmark{h}  \\ 
& AFGL5142-EC & 20.7(0.1) & --17.21(0.01) & 3.42(0.01) & 0.51(0.01) & 44.1(0.1)\tablefootmark{b} &           & 0.48(0.03) & --17.73(0.04) & 1.35(0.1) & 4.8(0.5) & 3.8(0.6) &   0.08(0.01) & 25\tablefootmark{h} \\   
&  05358-mm3 & 43.0(0.1) & --30.34(0.01) & 2.200(0.003) & 5(2) & 34(4)\tablefootmark{b} &                          & 0.16(0.02) & --30.5(0.2) & 2.9(0.4) & 9.3(0.9)          & 1.1(0.3) &  0.012(0.003) &  30\tablefootmark{h} \\ 
& G034-G2(MM2) & 2.0(0.1) & 27.2(0.02) &  1.950(0.003) & 1.51(0.01) & 7.57(0.03)\tablefootmark{b}         & & 0.76(0.03) & 26.80(0.08) & 1.6(0.3) & 1.7(0.2)      &  13(2) &   0.7(0.2) & $ - $ \\ 
 & G034-F1(MM8) & 1.65(0.05)  & 43.3(0.1) & 2.84(0.3) & $<$0.1 & 16\tablefootmark{c} &                         & 0.25(0.03) & 42.7(0.2) & 1.4(0.9) & 0.42(0.04)              &  1.8(0.3) &    0.43(0.09)  & $-$  \\ 
& G034-F2(MM7)  & 1.08(0.08) & 43.1(0.2) & 2.5(0.5) & -- & 16\tablefootmark{c} &                                       & 0.17(0.03) & 42(4) & 2.0(1.5) & 0.28(0.03)                    & 1.2(0.3) &      0.4(0.1) &  $-$ \\ 
& G028-C1(MM9)  & 2.4(0.1) & 65.32(0.03) & 2.4(0.2) & 3(1) & 6.4(0.4)\tablefootmark{b} &                       & 0.50(0.03) & 65.10(0.08) & 2.6(0.3) &  3.4(0.3)             & 13(2) &   0.38(0.07)  &  17\tablefootmark{i}  \\ 
& I20293-WC & 9.4(0.1) & --7.510(0.008) & 3.560(0.003) & $<$0.1 & 8.5\tablefootmark{a}  &                        & 0.82(0.03) & --7.7(0.3) & 1.0(0.9) & 5.9(0.6)              & 11(2) &   0.19(0.3) &  17\tablefootmark{h}  \\ 
& I22134-G & 6.9(0.1) & --33.20(0.01) & 1.01(0.02) & 3.3(0.2) & 15.9(0.3)\tablefootmark{b} &                 & 0.06(0.01) & --34.7(0.2) & 0.8(0.5) & 1.8(0.2)                 & 0.4(0.2) &    0.023(0.009) &  25\tablefootmark{h} \\ 
& I22134-B & 2.4(0.1) & --33.3(0.04) & 0.83(0.03) & 2.3(0.3) & 10.4(0.4)\tablefootmark{b} &                 & 0.09(0.02) & --33.8(0.06) & 0.5(0.4) &  1.0(0.1)               & 0.9(0.2) &    0.09(0.02) &  17\tablefootmark{h} \\ 
\cline{1-15}
HMPOs & I00117-MM1 & 5.5(0.1) & --50.89(0.01) & 1.50(0.04) & 2.4(0.3) & 19.3(0.4)\tablefootmark{b} &          & $\leq 0.07$ & -- & $--$                      & 1.26(0.1) &   $\leq 0.5$ & $\leq 0.04$ &  20\tablefootmark{h}  \\ 
& I04579-VLA1\tablefootmark{\dagger\dagger\dagger}  & 1.23(0.02) & --32.51(0.06) & 1.8(0.2) & $<$0.1 & 74\tablefootmark{d}  & -- & $\leq 0.14$ & & $--$ & 0.4(0.05) &   $\leq 1.5$ & $\leq 0.4$ &  74\tablefootmark{j}  \\ 
& AFGL5142-MM & 27.2(0.1) & --17.470(0.004) & 2.820(0.006) & 3.4(0.1) & 39.24(0.01)\tablefootmark{b} & & 0.40(0.02) & --17.7(0.1) & 2.3(0.3)      & 6.1(0.6) &  3.0(0.5) &   0.049(0.009) &  34\tablefootmark{h}  \\
& 05358-mm1 & 30.3(0.1) & --30.588(0.003) & 2.040(0.003) & 5(1) & 46(2)\tablefootmark{b} &           & 0.15(0.02) & --30.7(0.3) & 2.5(0.6)                   & 7.2(0.7) &   1.2(0.2) &  0.017(0.004)  & 39\tablefootmark{h} \\
& 18089--1732 & 31.4(0.2) & 17.40(0.004) & 4.960(0.001) & $<$0.1 & 38\tablefootmark{d}  &                       & 0.29(0.03) & 18.3(0.6) & 3(1)                  & 7.0(0.7) &   2.2(0.3) &   0.031(0.006)  &  38\tablefootmark{k}  \\ 
& 18517+0437 & 19.2(0.1) & 29.44(0.03) & 3.08(0.01) & $<$0.1 & 47\tablefootmark{e} &                             & 0.15(0.02) & 29.2(0.5) & 3(1)                   &  4.6(0.5) &   1.2(0.2) &  0.026(0.006)  & $ -$ \\ 
& G75-core & 6.5(0.1) & --14.99(0.02) & 3.99(0.04) & $<$0.1  & 96\tablefootmark{d}  &                                   & $\leq 0.05$ & -- & $--$                            & 2.4(0.2) &  $\leq 0.69$ &  $\leq 0.029$ &  96\tablefootmark{h} \\ 
& I20293-MM1 & 29.6(0.1) & --8.31(0.02) & 1.610(0.002) & 6.4(0.1) & 50.1(0.1)\tablefootmark{b} &   & 0.64(0.03) & --9.31(0.08) & 1.8(0.3)                &  7.3(0.7) &   5.4(0.8) &   0.07(0.01)  & 37\tablefootmark{h}  \\ 
& I21307  & 6.5(0.1) & --61.29(0.02) & 2.98(0.05) & $<$0.1 & 21\tablefootmark{d} &                                          & $\leq 0.06$ & -- & $--$                           & 1.5(0.1) &   $\leq 0.4$   & $\leq 0.03$ &  21\tablefootmark{j}  \\ 
& I23385 & 9.41(0.08) & --64.88(0.01) & 3.02(0.03) & $<$0.1 & 37\tablefootmark{d}  &                                 & 0.08(0.05) & --63(3) & 1.2(0.9)                &  2.1(0.2) &   0.6(0.1) &  0.028(0.009) &  37\tablefootmark{l}  \\ 
\cline{1-15}
UC HII & G5.89--0.39  & 84.1(0.3) & --5.820(0.003) & 5.900(0.009) & $<$0.1 & 20 \tablefootmark{f}  &      & 0.50(0.03) & --7.4(0.1) & 1.5(0.3)        &  19(2) &  3.4(0.5)  &  0.018(0.003) & $-$ \\ 
 & I19035-VLA1  & 13.7(0.2) & 17.62(0.01) & 4.49(0.03) & $<$0.1 & 19.0(0.2)\tablefootmark{g} &            & 0.18(0.03) & 17.5(0.1) & 1.4(0.3)          & 3.2(0.3) &   1.3(0.3) & 0.04(0.01) & 39\tablefootmark{h}  \\ 
& 19410+2336  & 33.3(0.1) & 7.90(0.005) &  3.550(0.001) & $<$0.1 & 21\tablefootmark{d} &                     & 0.52(0.05) & 7.5(0.08) & 1.5(0.3)         & 7.4(0.7) &  3.5(0.5)  & 0.047(0.009)  & 21\tablefootmark{k}   \\ 
& ON1  & 30.9(0.2) & --2.421(0.001) & 4.340(0.001) & $<$0.1 & 26\tablefootmark{d} &                                    & 0.17(0.05) & --3.3(0.9) & 3.13(1.65) &  6.6(0.7)   & 1.1(0.2) & 0.017(0.004) & 26\tablefootmark{m}  \\ 
& I22134-VLA1  & 2.5(0.08) & --33.11(0.01) & 1.30(0.06) & 1.9(0.4) & 10.6(0.5)\tablefootmark{b} &            &  0.08(0.01) & --34.2(0.4) & 0.7(1.0)   & 1.0(0.1) &   0.8(0.2) & 0.08(0.02) &  47\tablefootmark{h}  \\ 
& 23033+5951  & 19.6(0.1)  & --67.72(0.03) & 3.530(0.002) & $<$0.1 & 25\tablefootmark{d} &                 & 0.52(0.02) & --68.2(0.5) & 0.8(0.5)      & 4.4(0.4) &   3.5(0.5)  & 0.08(0.02) & 25\tablefootmark{k}   \\
& NGC7538-IRS9 & 13.9(0.1) & --71.814(0.006) & 3.42(0.03) & $<$0.1 & 20\tablefootmark{f} &                   & 0.14(0.02) & --72(2) & 2.6(0.5)        & 3.2(0.3) &   0.9(0.2) &   0.030(0.008) &  $ -$ \\ 
\hline
\end{tabular}
\end{center}
\tablefoot{
\tablefoottext{\dagger}{\Tex\ associated with uncertainties are calculated from the data, the others are assumed as explained in the footnotes;}
\tablefoottext{\dagger\dagger}{\Tk\ are derived from ammonia rotation temperatures following Tafalla et al.~\citeyear{tafalla};}
\tablefoottext{a}{\Tex$ \sim $1/2 \Tk , based on the results derived in this work for the HMSCs with well constrained \Tex\ and \Tk , except the "warm" ones;}
\tablefoottext{b}{derived from the (3--2) line parameters obtained from the hyperfine fitting procedure described in Sect.~\ref{obs}
and corrected for filling factor (see also the CLASS User Manual (http://iram.fr/IRAMFR/GILDAS/doc/html/class-html/class.html);}
\tablefoottext{c}{average value for HMSCs;}
\tablefoottext{d}{\Tex$ \sim $ \Tk , based on the results derived in this work for the HMPOs with well constrained \Tex\ and \Tk , and on the findings of Fontani et al.~(\citeyear{fontani06})
towards a sample of massive cores containing both HMPOs and UC HII regions;}
\tablefoottext{e}{average value for HMPOs;}
\tablefoottext{f}{average value for UCHIIs;}
\tablefoottext{g}{computed from the \H\ (1--0) line;}
\tablefoottext{h}{from VLA observations (Busquet~\citeyear{busquetphd},
S\'anchez-Monge~\citeyear{sanchez11}, Palau personal comm.); }
\tablefoottext{i}{from Effelsberg  observations (Pillai et al.~2006);}
\tablefoottext{j}{from Effelsberg  observations (Molinari et al.~1996);}
\tablefoottext{k}{from Effelsberg  observations (Sridharan et al~\citeyear{sridharan});}
\tablefoottext{l}{from VLA observations (Fontani et al.~2004b);}
\tablefoottext{m}{from Effelsberg observations (Jijina et al.~1999);}
\tablefoottext{\dagger\dagger\dagger}{\H\ parameters are derived from the (1--0) line, for which $\int T_{\rm MB}{\rm d}v$ is the integrated
intensity of the isolated component ($F_1 F = 1,0-1,1$) appropriately normalised and assumed to be optically thin. 
However, because observations were obtained under very bad weather conditions, the 
\Dfrac\ upper limit derived from this source is unreliable and we
decided to exclude it in the following analysis. }
}
\end{sidewaystable*}

\normalsize

\renewcommand{\thefigure}{B-\arabic{figure}}
\setcounter{figure}{0}
\section*{Appendix B: Spectra}
\label{appb}

In this appendix, all spectra of \D\ (2--1) and \H\ (3--2) transitions, for the sources
detected in both transitions, are shown.

\begin{figure*}
\centerline{
                     \includegraphics[angle=0,width=16cm]{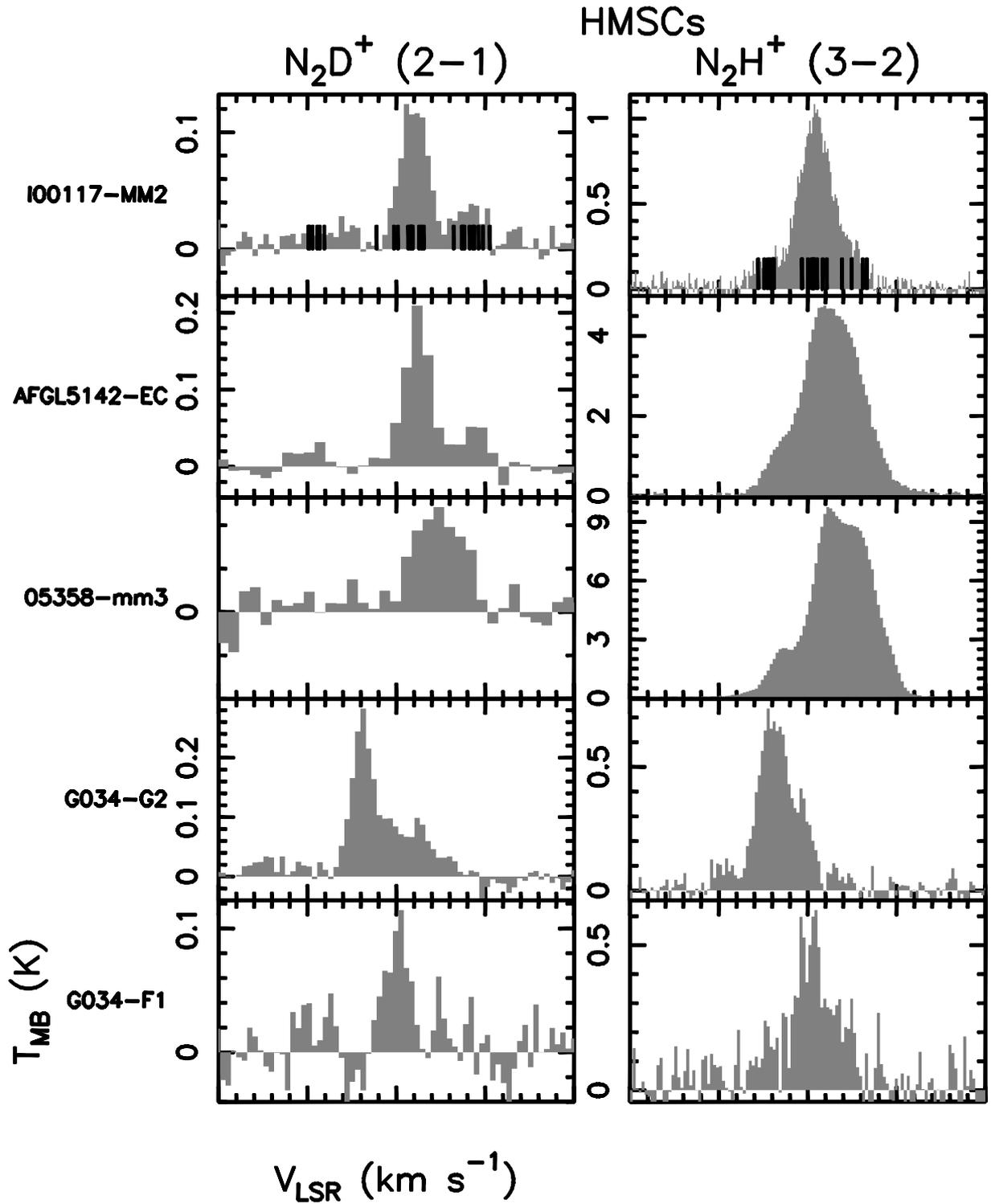}}
\caption[]{Spectra of \D\ (2--1) and 
\H\ (3--2) obtained towards the sources classified as HMSCs. 
For each spectrum, the velocity interval shown is $\pm 10$ \kms\ from the
systemic velocity listed in Table~\ref{tab_sources}.
The y-axis is in main beam brightness temperature units. In the 
spectra of I00117-MM2, the vertical bars show the position of the hyperfine 
components.}
\label{spectra1}
\end{figure*}

\begin{figure*}
\centerline{
                     \includegraphics[angle=0,width=16cm]{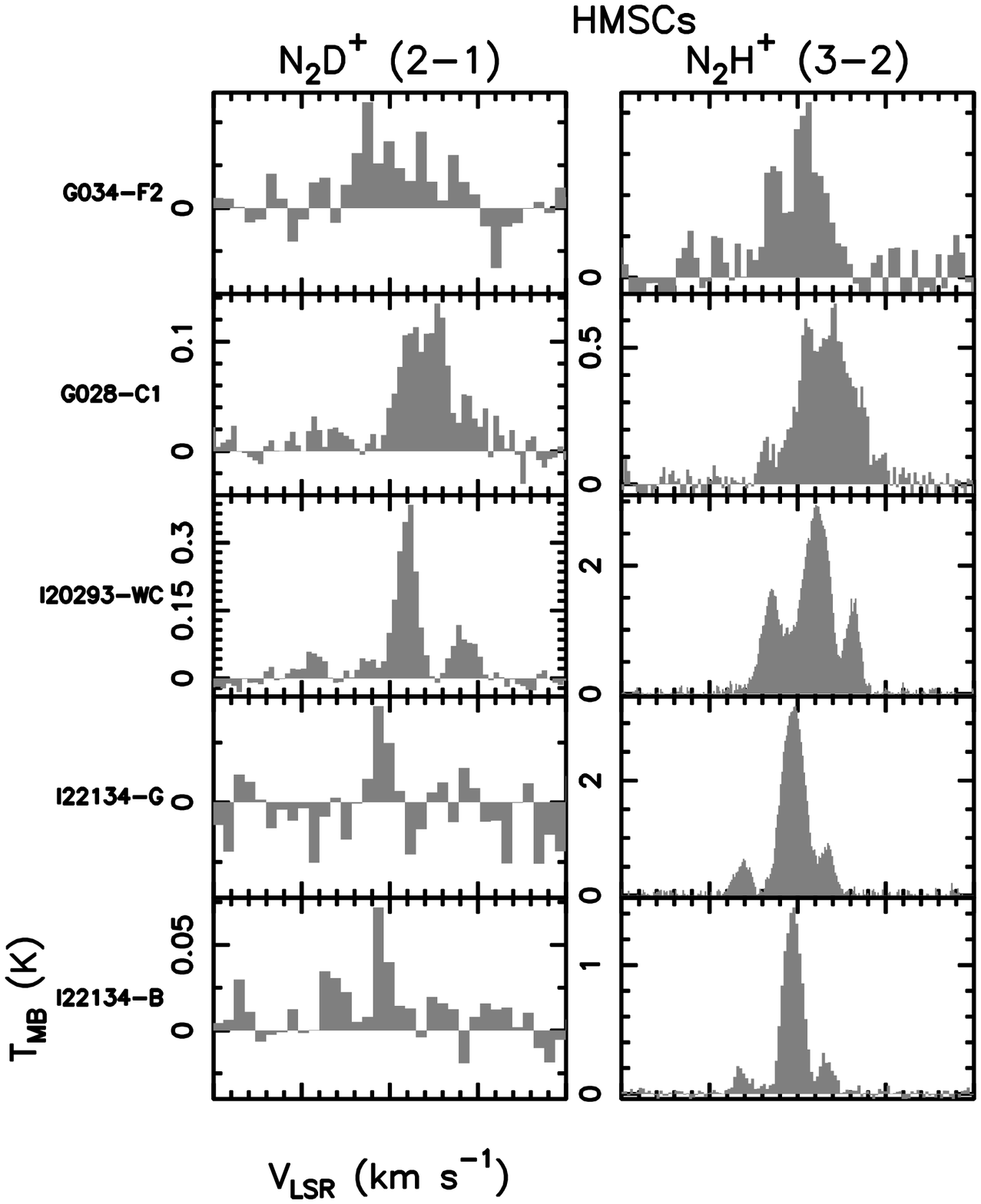}}
\caption[]{Fig.~\ref{spectra1} continued. }
\label{spectra2}
\end{figure*}

\begin{figure*}
\centerline{
                     \includegraphics[angle=0,width=16cm]{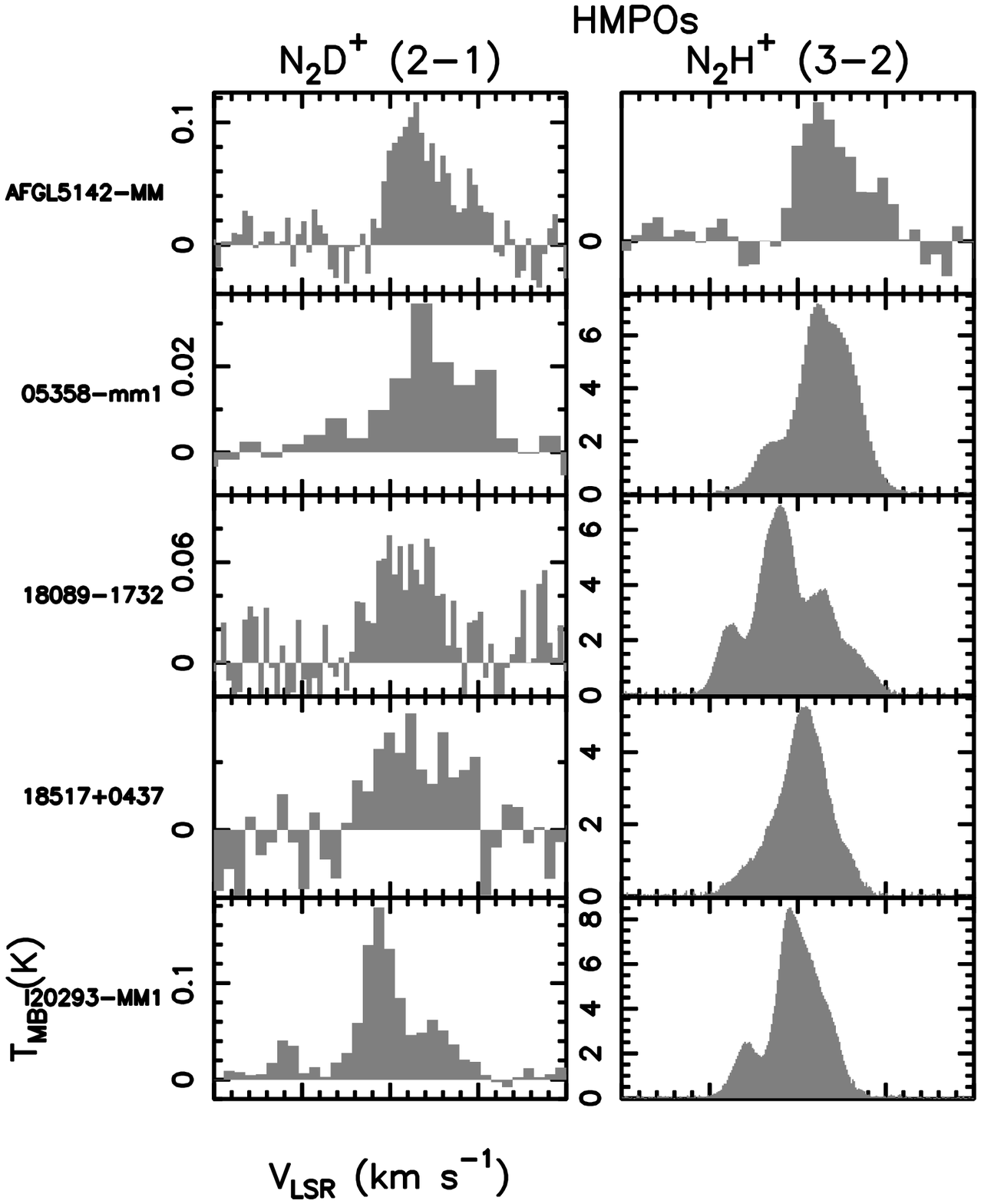}}
\caption[]{Same as Fig~\ref{spectra1} for HMPOs. }
\label{spectra3}
\end{figure*}

\begin{figure*}
\centerline{
                     \includegraphics[angle=0,width=16cm]{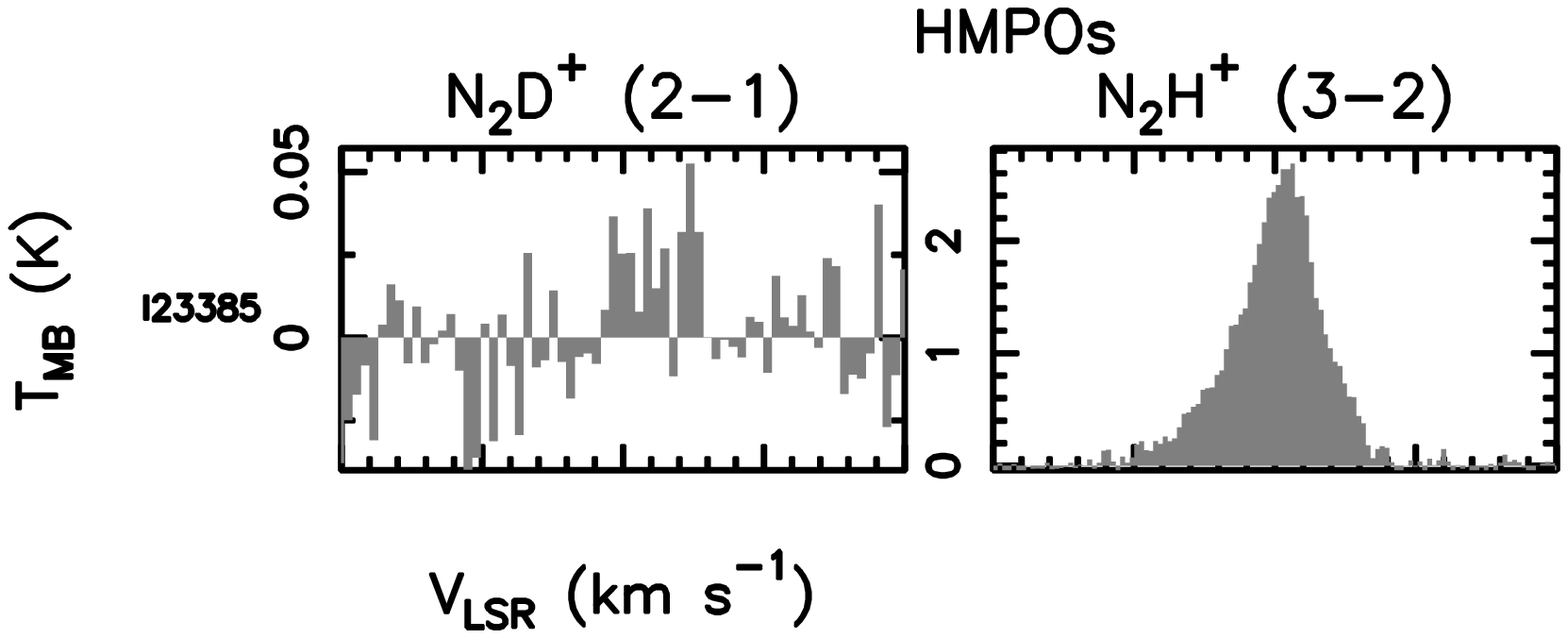}}
\caption[]{Fig.~\ref{spectra3} continued. }
\label{spectra4}
\end{figure*}

\begin{figure*}
\centerline{
                     \includegraphics[angle=0,width=16cm]{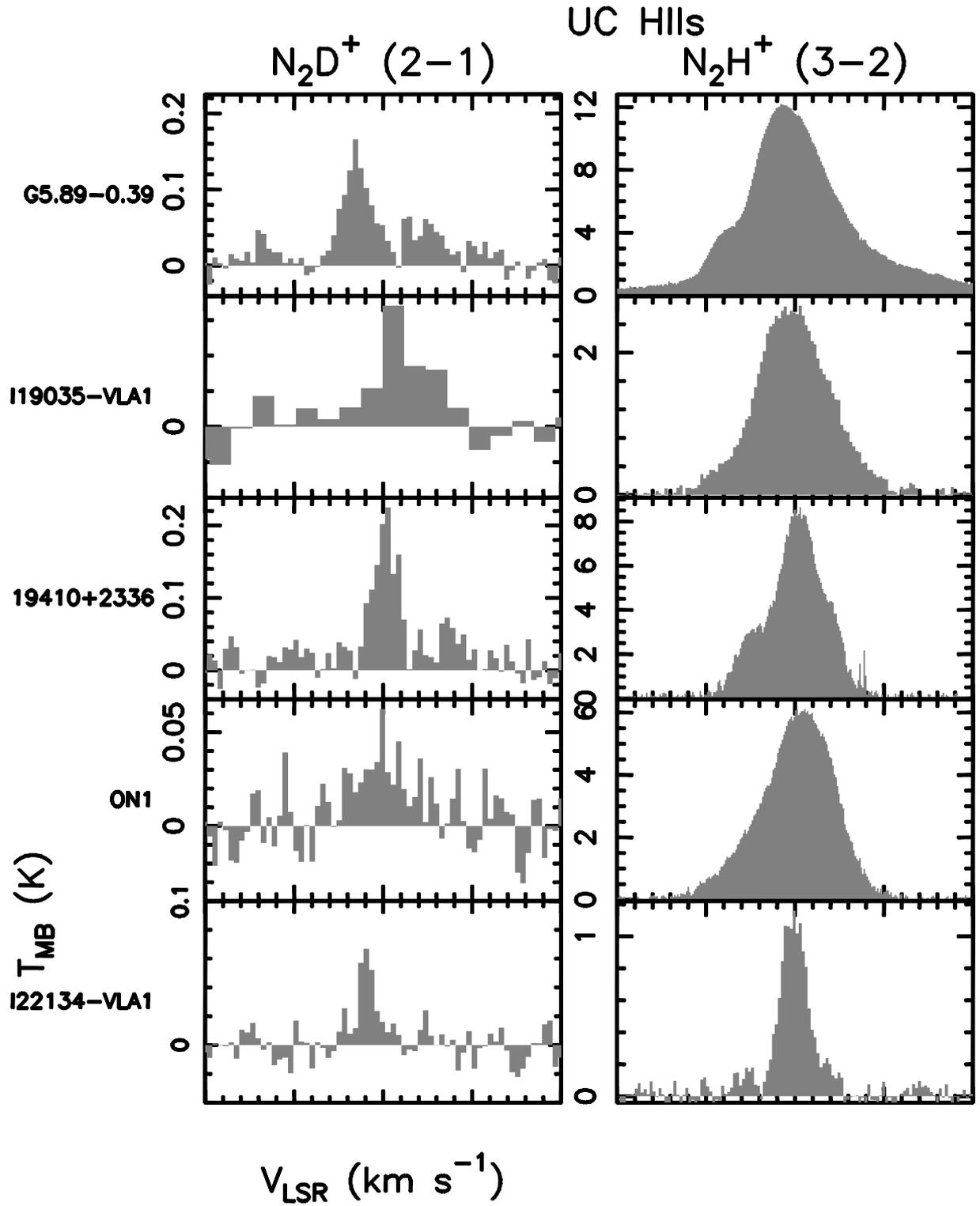}}
\caption[]{Same as Fig.~\ref{spectra1} for UC HII regions. }
\label{spectra5}
\end{figure*}

\begin{figure*}
\centerline{
                     \includegraphics[angle=0,width=16cm]{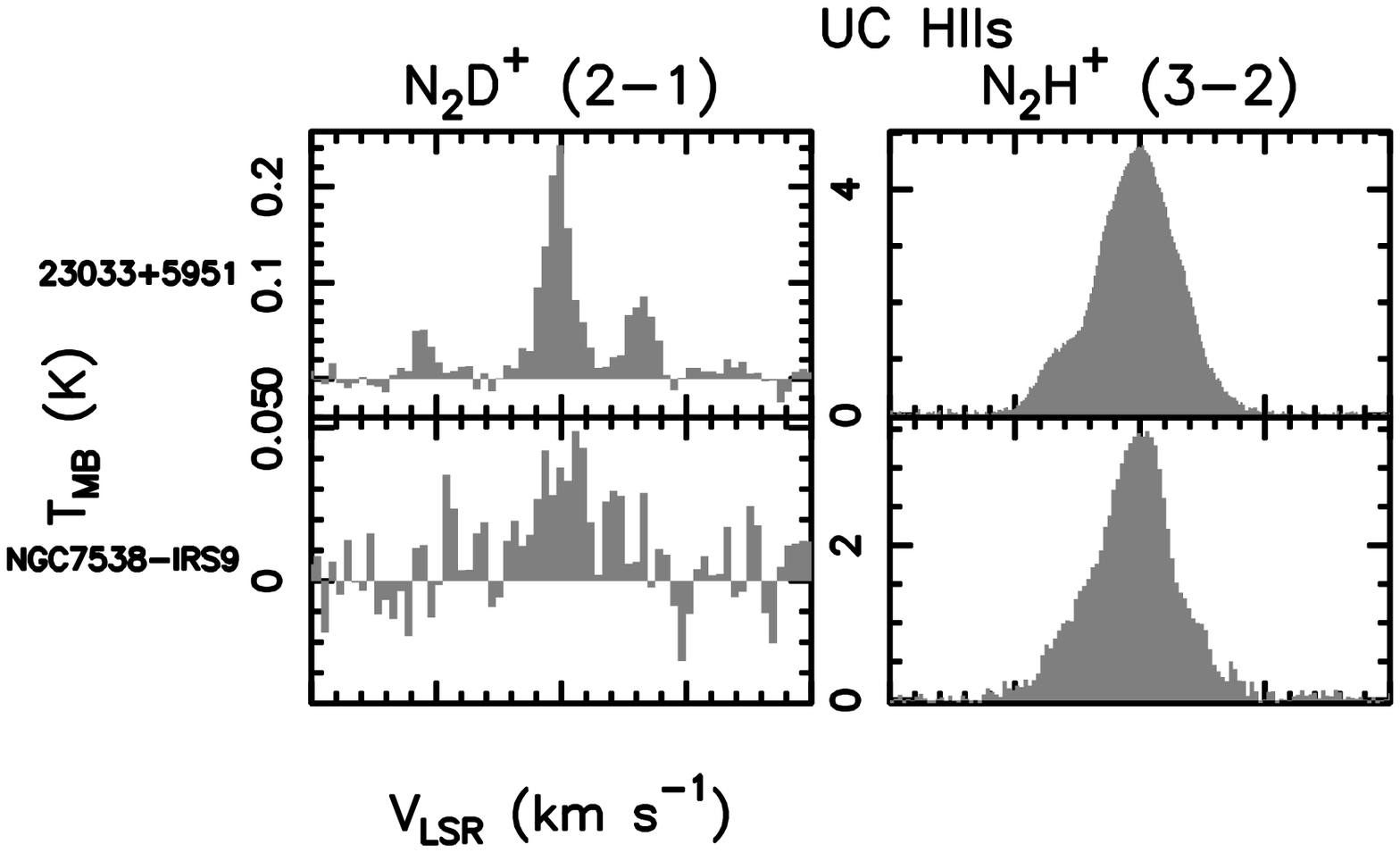}}
\caption[]{Fig.~\ref{spectra5} continued. }
\label{spectra6}
\end{figure*}
  
 \end{document}